\documentclass{article}

\usepackage[main,final]{neurips_2026}

\usepackage[utf8]{inputenc} 
\usepackage[T1]{fontenc}    
\usepackage{hyperref}       
\usepackage{url}            
\usepackage{booktabs}       
\usepackage{graphicx}       
\usepackage{amsfonts}       
\usepackage{nicefrac}       
\usepackage{microtype}      
\usepackage{xcolor}         
\usepackage[most]{tcolorbox} 

\title{Natural Language based Specification and Verification}

\author{%
  Zhaorui Li
  \quad Chengyu Song\thanks{Corresponding authors: Chengyu Song (csong@cs.ucr.edu) } \\
  University of California Riverside \\
}

\newtcolorbox{promptbox}[1]{
  enhanced,
  breakable,
  boxrule = 1.5pt,
  fontupper = \small,
  fonttitle = \bf\color{white},
  arc = 5pt,
  rounded corners,
  colframe=blue!50!green,
  colback=blue!5!white,
  coltitle=white,
  title = #1,
  left=4pt 
}

\begin{document}

\maketitle

\begin{abstract}
  Recent frontier large language models (LLMs) have shown strong performance in identifying security vulnerabilities in large, mature open-source systems. As LLM-generated code becomes increasingly common, a natural goal is to prevent such models from producing vulnerable implementations in the first place. Formal verification offers a principled route to this objective, but existing verification pipelines typically require specifications written in rigid formal languages. Prior work has explored using LLMs to synthesize such specifications, with limited success. In this paper, we investigate a different approach: using LLMs both to generate specifications and to verify implementations compositionally when the specifications are expressed in natural language. Our preliminary results suggest that this approach is promising.
\end{abstract}

\section{Introduction}

Memory safety violations remain a major source of software vulnerabilities in C and C++ programs. Errors such as invalid dereferences, use-after-free, double free, invalid free, and out-of-bounds accesses often arise from subtle interactions among allocation, deallocation, initialization, aliasing, and control flow. In realistic code, these properties are rarely confined to a single statement or even a single function: whether an operation is safe often depends on facts established in one procedure, propagated through several callees, and discharged only later in the calling context. This interprocedural structure makes memory-safety verification difficult for traditional tools. As a result, new memory errors keeps being introduced to codebases, including mature projects maintained by corporations with plenty of resources (e.g., the Google Chrome browser). Note that this challenge is not just for human developers and traditional tools, LLM-based coding agents also face the same set of challenges: lacking enough context, agents can introduce new memory errors too.

Recent progress in LLM-based bug finding makes this problem more urgent. LLMs are increasingly used not just to explain code, but to find bugs in real systems at a scale that was difficult to imagine with earlier tooling. In practice, LLM-assisted analysis can surface very large numbers of candidate bugs---for example, Linux kernel maintainers report that AI-generated bug reports increased from only two or three per week to roughly five to ten per day within about two years, with enough of those reports proving correct that additional maintainers were needed to handle them. This trend raises a broader question: once many bugs have been surfaced, how many bugs are still left in a code repository---or are there any left at all? And more critically, how can we prevent new bugs being introduced in the future? Answering these questions requires moving beyond bug finding, which seeks evidence of a feasible defect, to reasoning about whether defects are absent.

Formal verification offers one of the strongest ways to answer this question, because it aims not merely to find individual counterexamples, but to establish explicit safety guarantees. Its effectiveness, however, depends critically on having the right specifications. Automated verifiers do not prove vague intentions; they prove explicit claims. As a result, users often must provide preconditions, postconditions, loop invariants, frame conditions, and auxiliary assertions that capture intended behavior. In practice, this annotation burden remains a central obstacle to adoption. Decades of work on annotation assistants and invariant inference---including systems such as Houdini, Daikon, and Liquid Types---show both the value of automation and the difficulty of recovering rich semantic intent from syntax, traces, or restricted logical templates alone~\cite{flanagan2001houdini,ernst2007daikon,rondon2008liquid}. This makes formal verification a natural lens for assessing whether LLMs can move beyond surfacing bugs to reasoning about whether any bugs remain.

Recent LLM-based verification systems have renewed optimism that reducing this burden might become substantially easier. Most existing work uses LLMs primarily to generate formal proof artifacts like loop invariants, preconditions, and postconditions, that are then checked by external verifiers or theorem provers~\cite{pei2023invariants,chakraborty2023ranking,kamath2024leveraging,endres2024nl2postcond,wen2024autospec,wu2024lemur,ma2025specgen,banerjee2026dafnypro,mugnier2025laurel,silva2025daisy}. This design has yielded promising results for invariant generation and specification synthesis, but it places the LLM in a largely supportive role: the model proposes candidate formal objects, while the core reasoning and validation are delegated to a symbolic backend. Our work explores a different point in the design space. Instead of asking whether LLMs can help produce formal specifications for another verifier, we ask whether LLMs can themselves reason about memory safety more effectively when that reasoning is structured appropriately.

A straightforward way to pose this problem is to present an entire program to an LLM and ask whether it is memory-safe. However, this monolithic formulation is brittle. The model must, within a very long context, simultaneously infer the behavior of individual functions, recover how safety-relevant information flows across calls, and determine whether a violation is feasible under the accumulated calling context. Such long-context reasoning remain a major challenge for modern LLMs~\cite{bai2025longbench}. Whole-program prompts also obscure the modular structure that makes interprocedural verification tractable in the first place, forcing the model to rediscover local semantic facts each time it analyzes a new caller. Recent benchmarking results in formal verification point to the same broader limitation: local success on single functions does not readily translate into strong performance on multi-function programs~\cite{xu2025dafnycomp}: models that achieve over $99\%$ syntax correctness and over $58\%$ verification on prior single-function benchmarks fall to $95.67\%$ syntax correctness and just $3.69\%$ verification on compositional tasks, a drop of roughly $92\%$. Even the strongest model reaches only about $7\%$ verification at Pass@8, and most models remain below $2\%$.

Our intuition is that LLM reasoning for memory safety should follow traditional wisdom of compositional veification---decomposed around reusable semantic summaries of functions. However, intead of translating intermediate results into rigid formal specifications for a downstream symbolic verifier, we keep them in natural language and feed them back to the LLM in later reasoning stages. This design is motivated by the observation that, because LLMs are trained on code, documentation, and natural language, natural-language summaries may be easier for the model to generate, interpret, and refine consistently than rigid formal notation. Natural language can also be more expressive than formal notations. 

Based on this intuition, we present NLForge, a compositional LLM-based framework for memory-safety verification of C/C++ programs. NLForge first extracts functions from the program and constructs a call graph. It then performs a sequence of bottom-up analysis passes, generating summaries for allocation behavior, deallocation behavior, initialization behavior, and memory-safety contracts. These summaries are propagated from callees to callers and reused in a final verification pass, allowing the model to reason about a function under explicitly stated assumptions about its callees rather than raw source code alone. This decomposition is especially important for memory-safety bugs, where the safety of an operation often depends on a chain of prior facts: whether a pointer was allocated, whether ownership was transferred, whether storage was initialized, whether an alias may have been freed, or whether a size relationship remains valid across calls.

We evaluate this design on a subsect of programs from the memory-safety category of the International Competition on Software Verification (SV-COMP) 2026 using several modern LLMs, and compare it against a non-compositional baseline that asks the model to analyze all relevant source in a single prompt. Our results show that compositional natural-language summaries can substantially improve performance on several benchmark families, particularly those that require stronger interprocedural and control-flow reasoning; and particularly for smaller and quantized models, which lack robust long-context reasoning capabilities. At the same time, the gains are not universal: some benchmark categories remain difficult, and in some cases decomposition introduces new failure modes. We view this not as a weakness of the study, but as one of its central findings. The question is not simply whether LLMs can verify memory safety, but under what representations and decompositions they become more reliable, and where their reasoning still breaks down.

More broadly, our work argues for a shift in how LLMs can be used in verification. Rather than treating them only as generators of formal artifacts for external proof engines, we show that they can also serve as reasoning agents over structured, reusable natural-language summaries. This opens a complementary direction for LLM-based verification research: designing intermediate representations that are not only semantically meaningful to humans and modular across procedures, but also well matched to the strengths and weaknesses of LLM reasoning itself.

Our contributions are as follows:
\begin{itemize}
    \item We present a compositional LLM-based framework for memory-safety verification of C/C++ programs that performs bottom-up interprocedural reasoning over the call graph.
    \item We introduce natural-language semantic summaries as an intermediate representation for LLM-based verification, in contrast to prior work that primarily generates formal specifications for downstream symbolic checkers.
    \item We design a staged analysis pipeline that decomposes memory-safety reasoning into reusable summaries of allocation, deallocation, initialization, and caller-side memory-safety obligations, and feeds these summaries back to the LLM during later reasoning stages.
    \item We provide an empirical comparison between compositional reasoning and a non-compositional whole-program prompting baseline on C/C++ memory-safety benchmarks, identifying both the settings where compositional summaries help and the failure modes that remain.
\end{itemize}
\section{Related Work}
\subsection{LLM for automated program verification}
Many recent works use LLMs to generate specifications that are then checked by automated symbolic verifiers. A large portion of these focuses on loop invariants, including direct invariant prediction, inductive invariant synthesis, and verifier-aware ranking of multiple invariant candidates~\cite{pei2023invariants,kamath2024leveraging,chakraborty2023ranking,faria2025loopinvariants}. More recent systems extend this idea to richer program expressiveness. \emph{nl2postcond,endres2024nl2postcond} translates natural-language intent into formal postconditions. \emph{AutoSpec}~\cite{wen2024autospec} generates preconditions, postconditions, and loop invariants, then uses a theorem prover to filter incorrect specifications. \emph{Lemur}~\cite{wu2024lemur} and \emph{DafnyPro}~\cite{banerjee2026dafnypro} further incorporate verifier feedback to repair failed specifications. \emph{SpecGen}~\cite{ma2025specgen} argues that raw verification failure information can be too abstract for LLMs and therefore introduces mutation operators to guide LLM refinement. While these works evaluate LLMs as components in automated verification pipelines, Sultan et al.~\cite{sultan2026llms} study LLMs' standalone ability to reason about termination problems. Our work similarly examines the standalone reasoning ability of LLMs, but focuses on memory-corruption problems in programs. 

\subsection{Memory-safety verification tools}
Classical software verifiers in SV-COMP typically reduce program correctness to automated reasoning over control-flow, data-flow, and logical constraints, combining techniques such as abstract interpretation, symbolic execution, satisfiability modulo theories (SMT) solving, interpolation, and counterexample-guided abstraction refinement (CEGAR)~\cite{clarke1996formal}. Early systems such as SLAM~\cite{ball2004slam}, BLAST~\cite{beyer2007software}, and CBMC~\cite{kroening2014cbmc} helped establish key verification paradigms, including predicate-abstraction-based CEGAR and bounded model checking; later SV-COMP tools evolved toward more modular and portfolio-style architectures. For example, CPAchecker~\cite{beyer2007configurable} provides a configurable framework in which different program analyses, such as value analysis, predicate abstraction, and k-induction, can be combined under a common configurable program analysis (CPA) formalism. Ultimate Automizer~\cite{sas09:trace-abstraction} advances automata-based and interpolation-driven verification through trace abstraction, refining infeasible error paths into inductive proofs. Symbiotic~\cite{slaby2013symbiotic} represents another direction, integrating program instrumentation, slicing, and symbolic execution to improve scalability on practical C programs. Together, these tools illustrate how modern verifiers have moved beyond single-technique model checking toward flexible combinations of complementary analyses tailored to diverse verification tasks.

\section{Natural-Language Compositional Verification}
\label{sec:problem}

\begin{figure}[t]
  \centering
  \includegraphics[width=\linewidth]{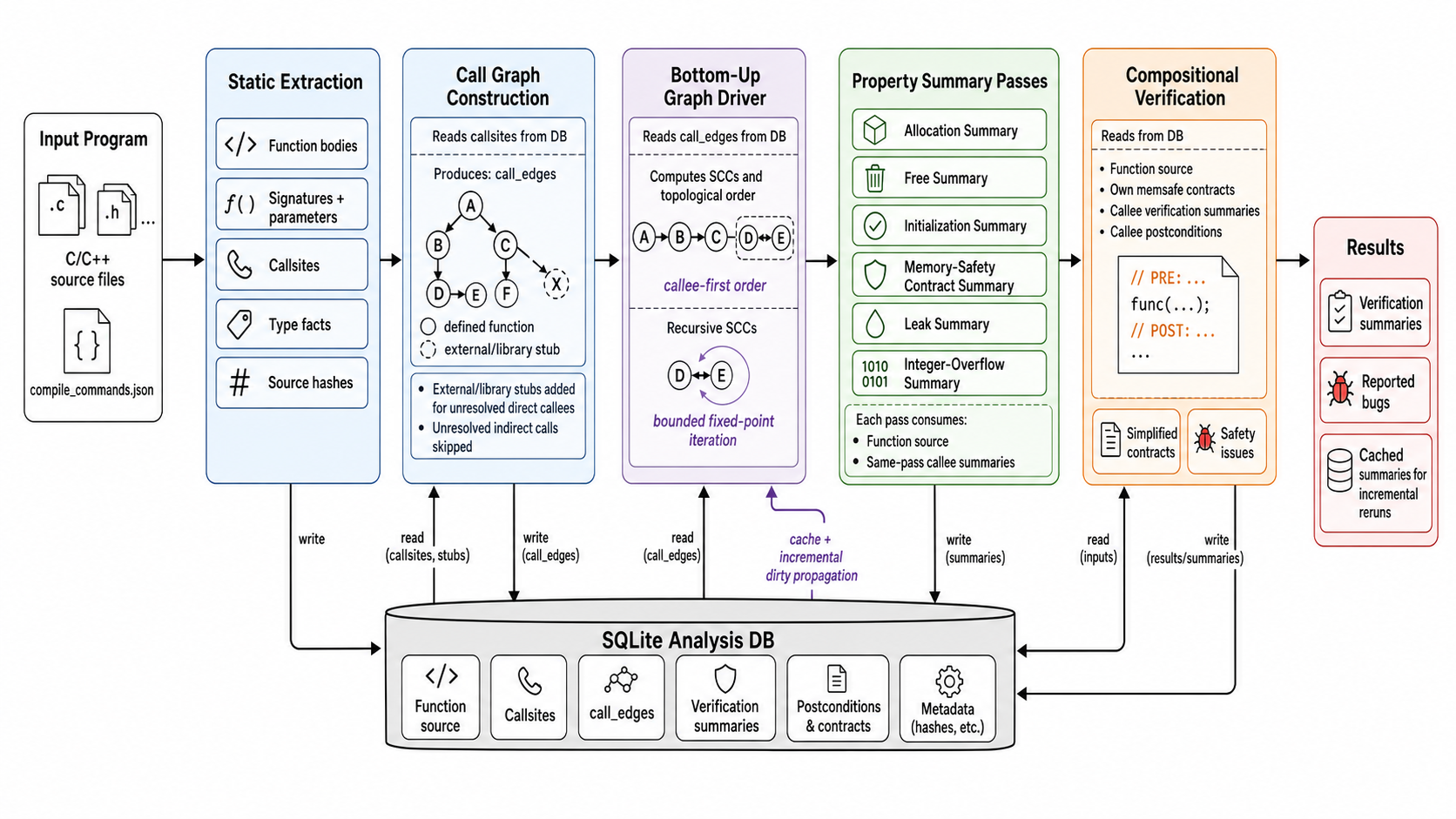}
  \caption{Overview of the NLForge pipeline. The system first extracts program structure and then performs bottom-up summary and verification passes over the call graph.}
  \label{fig:overview}
\end{figure}

We target the same verification problem as traditional modular safety
analysis: each function should be checked using a specification for the
functions it calls, and each function should expose a specification that its callers can use.  The difference is the representation and the source of these specifications.  Instead of requiring users to write formal contracts before running a verifier, NLForge infers structured natural-language specifications from the program and uses them to perform a bottom-up compositional check.

Figure~\ref{fig:overview} gives an overview of our analysis pipeline. The pipeline is designed around a simple compositional invariant: before a function is analyzed, summaries for its callees should already be available. The system therefore separates the front-end extraction of program facts from a sequence of bottom-up LLM analyses over the call graph. Each LLM pass produces a compact summary in natural language, which is stored in a database and reused by later passes.

\subsection{Deductive verification}

Traditional deductive verifiers reason about programs against explicit specifications rather than informal intent. A specification is a semantic summary of program behavior. At a high level, a precondition denotes the set of states from which it is valid to execute a command, while a postcondition denotes the set of states---or more precisely, the relation between entry and exit states---that the command may produce. Classic Hoare logic captures this idea with the triple
\[
  \{P\}\ C\ \{Q\},
\]
which states that if command $C$ starts in a state satisfying $P$, then every terminating execution ends in a state satisfying $Q$~\cite{hoare1969axiomatic}. 

The typical workflow of deductive verification is modular: for each function, the user writes a contract in terms of pre-conditions, post-conditions, and loop invariants. The verifier then translates the annotated program into verification conditions and checks them with a proof engine such as an SMT solver. When verifying a caller, the verifier does not repeatedly inline the full body of every callee; instead, it assumes that each callee satisfies its contract and checks whether the caller establishes the callee's preconditions before the call and uses the callee's postconditions after the call~\cite{hoare1969axiomatic,leino2010dafny}.

More formally, let a program be represented as a set of functions \(F\) and a directed call
graph \(G = (F, E)\), where an edge \((f,g) \in E\) means that \(f\) may call
\(g\).  For each function \(f\), our goal is to infer a compact boundary
specification \(S_f\) that is strong enough for callers to check their own
obligations, but small enough to avoid inlining the transitive call tree.  The
intended rule is Hoare-like:
\[
  \{P_f\}\ f\ \{Q_f\},
\]
where \(P_f\) is the set of caller obligations and \(Q_f\) is the set of
caller-visible effects.  A reported issue should correspond to a feasible
violation under \(P_f\).  Obligations that are merely required of callers are
kept as contracts rather than reported as bugs inside \(f\).

This decomposition is what makes traditional verification scale beyond single functions. A failed proof can mean that the implementation violates its specification, but it can also mean that the specification is missing a crucial assumption or guarantee. In practice, much of the effort in deductive verification therefore goes into writing the right preconditions, postconditions, loop invariants, assertions, and frame conditions: the proof engine can automate many logical steps, but it still depends on these annotations to expose the intended semantics of the program.




\subsection{Natural-Language Specifications}

NLForge represents \(S_f\) as a collection of typed JSON records. Fields such as \texttt{target}, \texttt{contract\_kind}, \texttt{size\_expr},
\texttt{condition}, and \texttt{severity} are compact summaries, while
\texttt{description} fields preserve natural-language explanations for facts. Summaries may also contain symbolic expressions expressed as code.
Overall, we only provide a JSON schema and high-level guidances in the prompt, LLMs have freedom to generate the contracts, rather than enforcing the use of a complete theorem-prover language.

For example, a memory-safety precondition can be represented as:
\begin{quote}
\small
\noindent
\texttt{\{ target: "p", contract\_kind: "not\_freed",
description: "p must not have been previously freed" \}}.
\end{quote}
A buffer-size obligation can record that \texttt{buf} must point to at least
\texttt{n} bytes, and an allocation postcondition can record that a callee
returns a heap object of size \texttt{n * sizeof(T)} that may be null.

The summaries are decomposed by property.  Allocation summaries describe heap
objects produced by a function, including size, nullability, and whether the
object is returned or stored through caller-visible state.  Free summaries
record deallocation effects and whether pointers are reset.  Initialization
summaries describe output parameters, reachable fields, return values, output
ranges, and non-returning behavior.  Memory-safety contracts describe
preconditions such as non-nullness, allowed nullness, not-freed requirements,
initialized-memory requirements, and buffer-size obligations.  Leak and integer
summaries record unresolved allocation/free obligations, range constraints, and
integer undefined behavior.  Verification summaries finally record the
simplified contracts still visible to callers and any issues (i.e., unsatisfied
safety contracts) found in the function body.

\subsection{NLForge pipeline}
\label{sec:pipeline}

\textbf{Preprocessing.}
Figure~\ref{fig:overview} gives an overview of the analysis pipeline.  The
front end first parses the target program using a compilation database.  For
each translation unit, it extracts function definitions, signatures, formal
parameters, source locations, function bodies, and callsite metadata.  When
preprocessing is enabled, macro-expanded source is recorded; otherwise the
original source body is used.  The extractor also records type declarations and
selected layout information, such as typedefs, struct definitions, field
offsets, and resolved \texttt{sizeof} values when available.

After extraction, NLForge constructs a call graph from the stored callsites.
Direct calls are resolved by function name.  If a callee is not defined in the
program, the implementation creates an external stub and attaches known library
attributes when available, such as \texttt{noreturn}.  The current evaluation
does not resolve unresolved function pointer callsites; those callsites are
skipped unless they have already been resolved into ordinary call edges.

\textbf{Bottom-Up Generation.}
All summary-generation passes share the same graph driver.  The driver computes
strongly connected components and processes them in callee-first topological
order, so a function is analyzed with the summaries of its direct callees
already available.  Recursive components are handled by bounded fixed-point
iteration: NLForge first summarizes all functions in the component, then reruns
the component when a summary changes.

Each LLM pass defines how to load a cached summary, build a prompt for one
function, parse the JSON response, and store the result.  The prompt contains
the function source, signature, source location, extracted type and macro
context when available, and a callee context assembled from the database.  For
large functions that cannot fit into the context windows of an LLM,
NLForge divides the function into several blocks, using natural syntax boundaries
like \texttt{switch} cases, \texttt{if/else} blocks, etc.
Blocks are processed according to the control-flow at source code level:
earlier blocks are summarized and replaces with summaries in compact comments.

The first passes infer caller-visible effects: allocation, free, and
initialization summaries.  When a function delegates work to a helper, the
callee summaries allow the model to decide whether the effect is hidden
internally or propagated through the current function's interface.  The
memory-safety pass then infers preconditions.
The model therefore solves a local propagation problem: which callee obligations are
discharged by the current function, and which remain obligations for callers.


\textbf{Compositional Checking.}
The final verifier pass checks each function under its own inferred
preconditions.  Its task is not to report those preconditions again, but to
(1) verify that the body is safe assuming the preconditions hold, and
(2) to simplify the obligations that remain visible to callers.
The verifier prompt (\autoref{sec:verifierprompt}) contains the function source,
the function's raw memory-safety contracts, simplified callee
preconditions, and callee postconditions from the allocation, free,
initialization, leak, and integer passes.

Conceptually, the LLM is asked to maintain a latent execution state;
starting from a function's own pre-conditions, the state must satisfy
the safety requirements of each statement, then the state will be
updated based on the processed statement.
For function calls, NLForge inserts callee facts as \texttt{PRE} and
\texttt{POST} comments with actual arguments substituted for formals when
possible.  Before the call, the verifier is asked to check whether
the current abstract state satisfies the callee's preconditions;
after the call, it updates the state using the callee's postconditions.
Because this pass is also run bottom-up, callers see only 
the simplified interface of each callee rather than
the callee's full implementation.

\textbf{Non-Compositional Checking.}
To reflect the performance of compositional checking, we also developed
a non-compositional baseline. In this mode, the LLM is given the same
memory safety verification task, but no summaries are generated nor consumed. Instead, the whole code is given to the LLM, with other auxiliary information like macros and type definitions.
This mode only works when the whole project (e.g., individual benchmarks
from SV-COMP) can fit into the context window of the target LLM.


\section{Experiments}
\label{sec:experiments}

We evaluate NLForge to answer two key research questions.
{\bf RQ1}: can model LLMs serve as verifiers for memory safety properties?
We answer how far a natural-language-based verifier can get on standard verification benchmarks when compared with mature static verifiers.
This comparison tests whether the summaries defined in Section~\ref{sec:problem} are informative enough to support realistic interprocedural reasoning.
{\bf RQ2}: can free-form natural language serve as specification to enable compositional verification?
Here, we isolate the effect of compositionality itself by comparing NLForge with a whole-program prompting baseline that uses the same model family but does not build callee-first summaries.
This comparison tests whether the pipeline in Section~\ref{sec:pipeline} improves LLM verification beyond simply showing the model more source code.

\subsection{Benchmarks, Baselines, and Metrics}

We evaluate on a subset of the SV-COMP 2026 C memory-safety benchmarks using three properties: (1) \texttt{valid-memsafety} (out-of-bound access, null-pointer dereference, use-after-free, invalid-free), which requires allocation/free/initialization/memsafety and verification passes; (2) \texttt{valid-memcleanup} (memory-leak), which requires the memleak pass; and (3) \texttt{no-overflow} (number/integer overflow), which requires the int-overflow pass.
The subset covers a broad mix of benchmark families, including categories \texttt{Juliet\_Test, Data Structure, ControlFlow}.
The \texttt{Data Structure} category consists of tasks that require reasoning about heap-allocated objects, linked lists, and arrays. The \texttt{ControlFlow} category consists of tasks whose correctness depends primarily on control-flow structure and integer variables.
This gives us coverage over both synthetic and real-world code. In particular, many \texttt{ControlFlow} tasks come from real-world programs such as \texttt{ntdrivers} and \texttt{openssl}, making this category generally more complex and challenging than the more templated subsets.
For NLForge, we instantiate the compositional pipeline with four LLM backbones: Claude Sonnet-4.6, GPT-5.4, Gemini-3.1-flash-lite, and a locally hosted Qwen3.5-27B-UD-Q4.
As traditional baselines, we report Symbiotic, CPAchecker, and UAutomizer, the top three tools in the SV-COMP 2026 C.MemSafety category; their results are taken from the official SV-COMP reports.

For each tool/property pair, we report \texttt{TP}, \texttt{FP}, \texttt{TN}, \texttt{FN}, and \texttt{UNK} as performance metrics.
The positive class corresponds to satisfying the queried property.
We also report accuracy, precision, and recall.
We treat \texttt{UNK} as a negative outcome for accuracy and recall.
Note for verification, recall (no FN) is more important than precision (FPs).
For SV-COMP, the score verifiers using the formula: \texttt{TP: +1, TN: +2, FP: -16, FN: -32, UNK: 0}.

\subsection{{\bf RQ1}: can model LLMs serve as verifiers for memory safety properties?}

Table~\ref{tab:statics-category-results} compares NLForge with established verification tools.
The results show that LLM-based analysis can serve as a useful memory-safety verifier in a practical, bug-finding sense, but not yet as a replacement for traditional sound verification. The most important difference is coverage. All LLM runs return a definite answer for every benchmark instance, whereas the traditional verifiers often avoid an incorrect answer by returning \texttt{UNK}.
Across the three benchmark families, Sonnet-4.6 reports 1262 true positives and only 4 false negatives, giving it the strongest bug-detection recall among the LLMs. However, this comes with 354 false positives. GPT-5.4 is more conservative, with fewer false positives (230) but many more missed positives (165 false negatives). Qwen3.5-27B-Q4 and Gemini-3.1-flash-lite fall between these two extremes, but both still make nontrivial numbers of false-positive and false-negative decisions.

The comparison with traditional verifiers highlights the tradeoff. CPAchecker produces no false positives or false negatives in the reported results, and Symbiotic makes only two false-negative decisions, but they leave 247 and 301 cases unknown, respectively. UAutomizer is even more conservative, with 895 unknown results. Thus, the LLMs are attractive when the goal is to obtain a complete classification or to prioritize likely memory-safety bugs, especially on benchmarks where conventional tools time out or fail to decide. They are less appropriate when the result must be interpreted as a formal proof of safety.

The category breakdown also matters. Juliet-style tests are the simplest setting: they are synthetic, relatively localized, and usually expose the target memory-safety pattern directly. In this category, the LLMs identify most positive instances, with Sonnet-4.6 matching CPAchecker on true positives, although still producing false alarms. Data-structure benchmarks are harder because they require reasoning about heap shape, ownership transfer, and aliasing from source text alone, and all LLMs produce many false positives in this category. The control-flow benchmarks are the most challenging despite their smaller count because they contain real-world programs with more complex path structure and less regular code. The best LLM results are close to the traditional tools on this category, but false positives and missed positives remain.

Overall, the answer to RQ1 is qualified. LLMs can act as high-coverage, high-recall memory-safety verifiers that are useful for bug finding and triage, but the observed false positives and false negatives mean that they cannot yet serve as standalone, sound verifiers for memory-safety properties.

\begin{table}[t]
\centering
\caption{Static analysis results grouped by benchmark category. TP, FP, TN, FN, and UNK are original counts. Tools marked with $\dagger$ are traditional verifiers.}
\label{tab:statics-category-results}
\begin{tabular}{llrrrrr}
\toprule
Category & Tool & TP & FP & TN & FN & UNK \\
\midrule
Juliet\_Test & Sonnet-4.6 & 921 & 98 & 830 & 0 & 0 \\
 & GPT-5.4 & 805 & 34 & 894 & 116 & 0 \\
 & Qwen3.5-27B-Q4 & 889 & 147 & 781 & 32 & 0 \\
 & Gemini-3.1-flash-lite & 894 & 186 & 742 & 27 & 0 \\
 & CPAchecker$^\dagger$ & 921 & 0 & 928 & 0 & 0 \\
 & Symbiotic$^\dagger$ & 921 & 0 & 913 & 0 & 15 \\
 & UAutomizer$^\dagger$ & 446 & 0 & 891 & 0 & 512 \\
\midrule
data structure & Sonnet-4.6 & 255 & 251 & 318 & 4 & 0 \\
 & GPT-5.4 & 223 & 196 & 373 & 36 & 0 \\
 & Qwen3.5-27B-Q4 & 215 & 182 & 387 & 44 & 0 \\
 & Gemini-3.1-flash-lite & 203 & 234 & 335 & 56 & 0 \\
 & CPAchecker$^\dagger$ & 168 & 0 & 414 & 0 & 246 \\
 & Symbiotic$^\dagger$ & 223 & 0 & 328 & 1 & 276 \\
 & UAutomizer$^\dagger$ & 91 & 0 & 360 & 0 & 377 \\
\midrule
control-flow & Sonnet-4.6 & 86 & 5 & 18 & 0 & 0 \\
 & Qwen3.5-27B-Q4 & 77 & 1 & 22 & 9 & 0 \\
 & GPT-5.4 & 73 & 0 & 23 & 13 & 0 \\
 & Gemini-3.1-flash-lite & 63 & 4 & 19 & 23 & 0 \\
 & CPAchecker$^\dagger$ & 85 & 0 & 23 & 0 & 1 \\
 & Symbiotic$^\dagger$ & 76 & 0 & 22 & 1 & 10 \\
 & UAutomizer$^\dagger$ & 83 & 0 & 20 & 0 & 6 \\
\bottomrule
\end{tabular}
\end{table}

\subsection{{\bf RQ2}: can free-form natural language serve as specification to enable compositional verification?}

Table~\ref{tab:compositional-category-results} compares the non-compositional baseline with the compositional NLForge pipeline, where caller-visible preconditions and effects are represented as natural-language specifications. Overall, the results suggest that natural language can serve as an effective intermediate specification format for compositional verification, although the gains depend on model and benchmark category.

The clearest benefit is better propagation of bug-relevant facts across function boundaries. Across all model/category pairs, the compositional runs increase true positives from 3390 to 3476 and reduce false negatives from 301 to 216. The strongest effect is for Qwen3.5-27B-Q4, whose true positives rise from 1013 to 1135 while false negatives fall from 218 to 97. The gains are especially visible on the data-structure benchmarks and the real-world control-flow benchmarks, suggesting that natural-language summaries help carry ownership, initialization, aliasing, and callee-precondition information from callees to callers without inlining the full call tree.

The control-flow category provides the strongest evidence for genuinely compositional reasoning. These benchmarks contain real-world programs and more complex path structure than Juliet or the data-structure microbenchmarks, and all three models reduce false negatives under composition: Sonnet-4.6 improves from 5 to 0, Qwen3.5-27B-Q4 from 44 to 9, and Gemini-3.1-flash-lite from 35 to 23. More broadly, composition appears most valuable on larger, more realistic tasks where whole-program reasoning exceeds the model's effective context or reasoning budget. On templated Juliet benchmarks, the benefit is less consistent because the non-compositional baseline can often exploit local bug patterns directly. Frontier models mainly gain on the hardest cases, whereas smaller models can receive a larger capability boost, but only when the generated summaries are reliable enough not to compound errors across calls.

Overall, the answer to RQ2 is yes, with qualifications. Free-form natural-language specifications are expressive enough to enable bottom-up compositional verification and can substantially improve recall when memory-safety facts must cross function boundaries, especially on large or path-sensitive tasks. At the same time, their informality makes precision model-dependent, so they are best viewed as a practical specification layer for LLM-guided verification rather than a substitute for formal contracts.

\begin{table}
\centering
\small
\caption{Compositional analysis compared with the non-compositional baseline by benchmark category. Juliet rows are left blank because those experiments are still running.}
\label{tab:compositional-category-results}
\begin{tabular}{lllrrrrr}
\toprule
Category & Model & Run & TP & FP & TN & FN & ERR \\
\midrule
Juliet\_Test & Sonnet-4.6 & Baseline & 897 & 153 & 661 & 0 \\
 &  & Comp. & 897 & 96 & 718 & 0 \\
 & Qwen3.5-27B-Q4 & Baseline & 853 & 63 & 751 & 44 \\
 &  & Comp. & 866 & 137 & 677 & 31 \\
 & Gemini-3.1-flash-lite & Baseline & 893 & 307 & 507 & 4 \\
 &  & Comp. & 872 & 183 & 631 & 25 \\
\midrule
data structure & Sonnet-4.6 & Baseline & 235 & 110 & 370 & 13 \\
 &  & Comp. & 243 & 206 & 273 & 6 \\
 & Qwen3.5-27B-Q4 & Baseline & 119 & 42 & 438 & 130 \\
 &  & Comp. & 192 & 139 & 341 & 57 \\
 & Gemini-3.1-flash-lite & Baseline & 223 & 176 & 304 & 26 \\
 &  & Comp. & 184 & 177 & 303 & 65 \\
\midrule
control-flow & Sonnet-4.6 & Baseline & 79 & 4 & 19 & 5 \\
 &  & Comp. & 82 & 5 & 18 & 0 \\
 & Qwen3.5-27B-Q4 & Baseline & 41 & 1 & 22 & 44 \\
 &  & Comp. & 77 & 1 & 22 & 9 \\
 & Gemini-3.1-flash-lite & Baseline & 50 & 7 & 16 & 35 \\
 &  & Comp. & 63 & 4 & 19 & 23 \\
\bottomrule
\end{tabular}
\end{table}

\section{Limitation and Future Work}

Our study has several limitations. We evaluate only a subset of SV-COMP 2026 memory-safety benchmarks and a small set of model backbones, so the results are not a comprehensive assessment of LLM-based verification across languages, properties, or tool settings. NLForge also depends on the quality of generated natural-language summaries: if summaries omit conditions, overgeneralize effects, or propagate mistakes, compositional reasoning can amplify rather than correct errors. In addition, the current pipeline does not fully address harder features such as unresolved indirect calls, richer heap structure, or the soundness guarantees expected from traditional formal verification.

A key direction for future work is to make compositional LLM-based verification a reusable analysis layer. Cached summaries could support incremental analysis in continuous integration, reducing both latency and token cost by re-verifying only affected procedures. The same decomposition may also help scale to larger codebases by replacing repeated whole-program prompting with compact summaries, even as longer context windows improve. Other promising directions include improving calibration, combining summaries with stronger symbolic or type-based checks, handling indirect calls and heap invariants more robustly, and extending the approach beyond memory safety.

\section{Conclusion}

We presented NLForge, a compositional verification framework that uses LLMs to infer natural-language summaries and verify memory-safety properties in a bottom-up, callee-first manner. By decomposing verification into reusable summaries, NLForge lets the model solve smaller local reasoning problems while still propagating caller-visible effects through the call graph. Across SV-COMP 2026 memory-safety benchmarks, the best NLForge configuration recovers much of the performance of mature static verifiers, especially on \texttt{valid-memsafety}, where Sonnet-4.6 reaches 89.6\% accuracy with 99.8\% recall. Its main weakness is calibration rather than a complete failure of interprocedural reasoning: compared with symbolic tools, NLForge produces far fewer unknown results but more false positives.

The compositional ablation shows that this structure helps most on harder, less templated tasks, where whole-program prompting is more likely to exceed the model's effective context or reasoning budget. For frontier models, compositionality yields incremental gains on difficult cases; for smaller models, it can be a stronger capability amplifier if the generated summaries are reliable. Overall, these results suggest that natural language can serve as a practical intermediate specification layer for compositional verification, while also highlighting the need for better calibration, more reliable summaries, and stronger support for challenging program features.

\bibliographystyle{plain}
\setlength{\bibsep}{3pt}
\bibliography{references}

@article{ernst2007daikon,
  title={The Daikon system for dynamic detection of likely invariants},
  author={Ernst, Michael D and Perkins, Jeff H and Guo, Philip J and McCamant, Stephen and Pacheco, Carlos and Tschantz, Matthew S and Xiao, Chen},
  journal={Science of computer programming},
  volume={69},
  number={1-3},
  pages={35--45},
  year={2007},
  publisher={Elsevier}
}

@inproceedings{flanagan2001houdini,
  title={Houdini, an annotation assistant for ESC/Java},
  author={Flanagan, Cormac and Leino, K Rustan M},
  booktitle={International symposium of formal methods Europe},
  pages={500--517},
  year={2001},
  organization={Springer}
}

@inproceedings{rondon2008liquid,
  title={Liquid types},
  author={Rondon, Patrick M and Kawaguci, Ming and Jhala, Ranjit},
  booktitle={Proceedings of the 29th ACM SIGPLAN Conference on Programming Language Design and Implementation},
  pages={159--169},
  year={2008}
}

@inproceedings{chakraborty2023ranking,
  title={Ranking llm-generated loop invariants for program verification},
  author={Chakraborty, Saikat and Lahiri, Shuvendu and Fakhoury, Sarah and Lal, Akash and Musuvathi, Madanlal and Rastogi, Aseem and Senthilnathan, Aditya and Sharma, Rahul and Swamy, Nikhil},
  booktitle={Findings of the Association for Computational Linguistics: EMNLP 2023},
  pages={9164--9175},
  year={2023}
}

@inproceedings{pei2023invariants,
  title={Can large language models reason about program invariants?},
  author={Pei, Kexin and Bieber, David and Shi, Kensen and Sutton, Charles and Yin, Pengcheng},
  booktitle={International Conference on Machine Learning},
  pages={27496--27520},
  year={2023},
  organization={PMLR}
}

@inproceedings{kamath2024leveraging,
  title={Leveraging llms for program verification},
  author={Kamath, Adharsh and Mohammed, Nausheen and Senthilnathan, Aditya and Chakraborty, Saikat and Deligiannis, Pantazis and Lahiri, Shuvendu K and Lal, Akash and Rastogi, Aseem and Roy, Subhajit and Sharma, Rahul},
  booktitle={2024 Formal Methods in Computer-Aided Design (FMCAD)},
  pages={107--118},
  year={2024},
  organization={IEEE}
}

@article{banerjee2026dafnypro,
  title={DafnyPro: LLM-Assisted Automated Verification for Dafny Programs},
  author={Banerjee, Debangshu and Bouissou, Olivier and Zetzsche, Stefan},
  journal={arXiv preprint arXiv:2601.05385},
  year={2026}
}

@inproceedings{wu2024lemur,
title={Lemur: Integrating Large Language Models in Automated Program Verification},
author={Haoze Wu and Clark Barrett and Nina Narodytska},
booktitle={The Twelfth International Conference on Learning Representations},
year={2024},
url={https://openreview.net/forum?id=Q3YaCghZNt}
}

@article{endres2024nl2postcond,
author = {Endres, Madeline and Fakhoury, Sarah and Chakraborty, Saikat and Lahiri, Shuvendu K.},
title = {Can Large Language Models Transform Natural Language Intent into Formal Method Postconditions?},
year = {2024},
issue_date = {July 2024},
publisher = {Association for Computing Machinery},
address = {New York, NY, USA},
volume = {1},
number = {FSE},
url = {https://doi.org/10.1145/3660791},
doi = {10.1145/3660791},
journal = {Proc. ACM Softw. Eng.},
month = jul,
articleno = {84},
numpages = {24}
}

@inproceedings{wen2024autospec,
  title={Enchanting program specification synthesis by large language models using static analysis and program verification},
  author={Wen, Cheng and Cao, Jialun and Su, Jie and Xu, Zhiwu and Qin, Shengchao and He, Mengda and Li, Haokun and Cheung, Shing-Chi and Tian, Cong},
  booktitle={International Conference on Computer Aided Verification},
  pages={302--328},
  year={2024},
  organization={Springer}
}

@inproceedings{ma2025specgen,
  title={Specgen: Automated generation of formal program specifications via large language models},
  author={Ma, Lezhi and Liu, Shangqing and Li, Yi and Xie, Xiaofei and Bu, Lei},
  booktitle={2025 IEEE/ACM 47th International Conference on Software Engineering (ICSE)},
  pages={16--28},
  year={2025},
  organization={IEEE}
}

@inproceedings{faria2025loopinvariants,
  title={Automatic generation of loop invariants in dafny with large language models},
  author={Pascoal Faria, Jo{\~a}o and Trigo, Emanuel and Abreu, Rui},
  booktitle={International Conference on Fundamentals of Software Engineering},
  pages={138--154},
  year={2025},
  organization={Springer}
}

@article{mugnier2025laurel,
  title={Laurel: Unblocking automated verification with large language models},
  author={Mugnier, Eric and Gonzalez, Emmanuel Anaya and Polikarpova, Nadia and Jhala, Ranjit and Yuanyuan, Zhou},
  journal={Proceedings of the ACM on Programming Languages},
  volume={9},
  number={OOPSLA1},
  pages={1519--1545},
  year={2025},
  publisher={ACM New York, NY, USA}
}

@misc{silva2025daisy,
  title={Inferring multiple helper Dafny assertions with LLMs},
  author={Silva, {\'A}lvaro and Mendes, Alexandra and Martins, Ruben},
  journal={arXiv preprint arXiv:2511.00125},
  year={2025}
}

@article{sultan2026llms,
  title={LLMs versus the Halting Problem: Revisiting Program Termination Prediction},
  author={Sultan, Oren and Armengol-Estape, Jordi and Kesseli, Pascal and Vanegue, Julien and Shahaf, Dafna and Adi, Yossi and O'Hearn, Peter},
  journal={arXiv preprint arXiv:2601.18987},
  year={2026}
}

@article{xu2025dafnycomp,
  title={Local Success Does Not Compose: Benchmarking Large Language Models for Compositional Formal Verification},
  author={Xu, Xu and Li, Xin and Qu, Xingwei and Fu, Jie and Yuan, Binhang},
  journal={arXiv preprint arXiv:2509.23061},
  year={2025}
}

@article{hoare1969axiomatic,
  title={An axiomatic basis for computer programming},
  author={Hoare, C. A. R.},
  journal={Communications of the ACM},
  volume={12},
  number={10},
  pages={576--580},
  year={1969},
  doi={10.1145/363235.363259}
}

@inproceedings{leino2010dafny,
  title={Dafny: An automatic program verifier for functional correctness},
  author={Leino, K. Rustan M.},
  booktitle={International Conference on Logic for Programming Artificial Intelligence and Reasoning},
  pages={348--370},
  year={2010},
  organization={Springer},
  doi={10.1007/978-3-642-17511-4\_20}
}

@inproceedings{ball2004slam,
  title={SLAM and Static Driver Verifier: Technology transfer of formal methods inside Microsoft},
  author={Ball, Thomas and Cook, Byron and Levin, Vladimir and Rajamani, Sriram K},
  booktitle={International Conference on Integrated Formal Methods},
  pages={1--20},
  year={2004},
  organization={Springer}
}

@misc{CPAchecker,
title = {{{\sc CPAchecker}}: Configurable Software Verification},
year = {2007},
url = {http://www.sosy-lab.org/~dbeyer/CPAchecker/},
role = {Principal designer, architect, implementation, and maintenance},
}

@inproceedings{beyer2007configurable,
  title={Configurable software verification: Concretizing the convergence of model checking and program analysis},
  author={Beyer, Dirk and Henzinger, Thomas A and Th{\'e}oduloz, Gr{\'e}gory},
  booktitle={International Conference on Computer Aided Verification},
  pages={504--518},
  year={2007},
  organization={Springer}
}

@inproceedings{sas09:trace-abstraction,
  author    = {Matthias Heizmann and Jochen Hoenicke and Andreas Podelski},
  editor    = {Jens Palsberg and Zhendong Su},
  title     = {Refinement of Trace Abstraction},
  booktitle = {Static Analysis, 16th International Symposium, {SAS} 2009, Los Angeles,
               CA, USA, August 9-11, 2009. Proceedings},
  series    = {Lecture Notes in Computer Science},
  volume    = {5673},
  pages     = {69--85},
  publisher = {Springer},
  year      = {2009},
  doi       = {10.1007/978-3-642-03237-0\_7},
}

@inproceedings{slaby2013symbiotic,
  title={Symbiotic: Synergy of Instrumentation, Slicing, and Symbolic Execution: (Competition Contribution)},
  author={Slaby, Jiri and Strej{\v{c}}ek, Jan and Trt{\'\i}k, Marek},
  booktitle={International Conference on Tools and Algorithms for the Construction and Analysis of Systems},
  pages={630--632},
  year={2013},
  organization={Springer}
}

@inproceedings{kroening2014cbmc,
  title={CBMC--C Bounded Model Checker: (Competition Contribution)},
  author={Kroening, Daniel and Tautschnig, Michael},
  booktitle={International Conference on Tools and Algorithms for the Construction and Analysis of Systems},
  pages={389--391},
  year={2014},
  organization={Springer}
}

@article{beyer2007software,
  title={The software model checker Blast: Applications to software engineering},
  author={Beyer, Dirk and Henzinger, Thomas A and Jhala, Ranjit and Majumdar, Rupak},
  journal={International Journal on Software Tools for Technology Transfer},
  volume={9},
  number={5},
  pages={505--525},
  year={2007},
  publisher={Springer}
}

@article{clarke1996formal,
  title={Formal methods: State of the art and future directions},
  author={Clarke, Edmund M and Wing, Jeannette M},
  journal={ACM Computing Surveys (CSUR)},
  volume={28},
  number={4},
  pages={626--643},
  year={1996},
  publisher={ACM New York, NY, USA}
}

@inproceedings{bai2025longbench,
  title={Longbench v2: Towards deeper understanding and reasoning on realistic long-context multitasks},
  author={Bai, Yushi and Tu, Shangqing and Zhang, Jiajie and Peng, Hao and Wang, Xiaozhi and Lv, Xin and Cao, Shulin and Xu, Jiazheng and Hou, Lei and Dong, Yuxiao and others},
  booktitle={63rd Annual Meeting of the Association for Computational Linguistics},
  pages={3639--3664},
  year={2025}
}


\appendix
\section{Broader Impact and Societal Implications}

Our work studies whether large language models can act as practical verifiers for memory-safety properties. We believe the main positive societal impact is improved software reliability. Because an LLM-based verifier is comparatively easy to prototype and extend, this line of work could help bring verification-style reasoning to settings where traditional formal methods remain too expensive or specialized to deploy. In the long run, such tools could help strengthen safety-critical software in domains such as medical devices and autonomous vehicles, where memory-safety failures can have serious real-world consequences. The same accessibility may also democratize formal methods for smaller or resource-constrained teams that cannot afford dedicated verification experts.

At the same time, the approach also carries meaningful risks. First, the same capability that helps defenders identify safety violations could be used by attackers to find bugs and exploit them for harmful purposes. Second, false negatives are a particularly important failure mode in verification: if an LLM-based verifier incorrectly labels an unsafe program as safe, developers may place unjustified trust in software that still contains dangerous flaws. For this reason, we view current LLM verifiers as decision-support tools rather than stand-alone guarantees of correctness.

There are also concerns around accessibility and resource usage. Our evaluation relies in part on frontier-model API access, which can be expensive and may create disparities between well-resourced organizations and smaller groups that want to use the same verification pipeline. In addition, extensive evaluation across four LLM backbones incurs nontrivial computational cost, and we acknowledge the associated environmental footprint. These concerns suggest that future work should emphasize efficient evaluation, stronger open and local alternatives, and careful deployment practices that avoid over-trusting model outputs.

\section{Licenses of Used Code}
We use existing benchmarks and verification tools in our evaluation, and we summarize their licenses here. The SV-COMP benchmark materials are distributed under \texttt{CC-BY-4.0} (SPDX identifier: \texttt{CC-BY-4.0}). CPAchecker is licensed under the Apache License 2.0. Symbiotic is licensed under the MIT License. UAutomizer is part of the Ultimate framework; the core of Ultimate and many of its plugins are licensed under LGPLv3 with a linking exception for Eclipse RCP and Eclipse CDT.

For the language models used in our experiments, Qwen3.5-27B was locally hosted and distributed as an open-weight model under the Apache License 2.0. By contrast, the other three models were accessed as proprietary hosted services rather than redistributed model weights. GPT-5.4 was used through the OpenAI API and is governed by OpenAI's service terms and services agreement. Claude Sonnet-4.6 was used through Anthropic's API and is governed by Anthropic's Commercial Terms of Service. Gemini-3.1-flash-lite was used through the Gemini API and is governed by the Google APIs Terms of Service together with the Gemini API Additional Terms of Service.

\section{NLForge Prompt Templates}
\label{sec:prompt}

This appendix records the prompt templates used by NLForge. In the implementation, each prompt is instantiated by substituting placeholders such as \texttt{\{source\}}, \texttt{\{name\}}, \texttt{\{signature\}}, \texttt{\{file\_path\}}, and \texttt{\{callee\_summaries\}}. Below we reproduce the allocation-summary prompt family in paper-friendly form.

\subsection{Allocation Summarizer}
All allocation prompt variants require the model to return JSON in the following format.
\begin{promptbox}{JSON response schema}
\begin{verbatim}
{
  "function": "<function_name>",
  "description": "One-sentence summary of what this function allocates",
  "parameters": {
    "param_name": {
      "role": "role description",
      "used_in_allocation": true|false
    }
  },
  "allocations": [
    {
      "type": "heap|static|parameter_derived|escaped_stack",
      "source": "allocator or provenance",
      "size_expr": "size expression or null",
      "size_params": ["parameter names affecting size"],
      "returned": true|false,
      "stored_to": "field/variable name or null",
      "may_be_null": true|false
    }
  ],
  "buffer_size_pairs": [
    {
      "buffer": "buffer variable/field",
      "size": "size variable/field",
      "kind": "param_pair|struct_field|flexible_array",
      "relationship": "byte count|element count|max capacity"
    }
  ]
}
\end{verbatim}
\end{promptbox}

\paragraph{Single-message prompt.}
The simplest variant places the function source, metadata, callee summaries, and the shared instruction block into a single prompt:

\begin{promptbox}{Single-message prompt.}
\begin{verbatim}

You are analyzing C/C++ code to generate memory allocation summaries.

## Function to Analyze
```c
{source}
```

Function: {name}
Signature: {signature}
File: {file_path}

## Callee Summaries
{callee_summaries}

## Task
Generate a memory allocation summary for this function.
[shared allocation instructions]
[JSON schema]
\end{verbatim}
\end{promptbox}

\paragraph{Approach A: cached instructions.}
In the \texttt{cache\_mode="instructions"} configuration, the static task instructions are placed in the system prompt and reused across functions, while the per-function source and callee summaries appear in the user message.
\begin{promptbox}{Approach A: cached instructions.}
\begin{verbatim}
System prompt:
You are analyzing C/C++ code to generate memory allocation summaries.
## Task
Generate a memory allocation summary for the function provided in the
user message.
[shared allocation instructions]
[JSON schema with <function_name> placeholder]

User prompt:
## Function to Analyze
```c
{source}
```
Function: {name}
Signature: {signature}
File: {file_path}
## Callee Summaries
{callee_summaries}
\end{verbatim}
\end{promptbox}

\paragraph{Approach B: cached source.}
In the \texttt{cache\_mode="source"} configuration, the function source is cached in the system prompt and the user prompt contains the task instructions and the callee summaries.
\begin{promptbox}{Approach B: cached source.}
\begin{verbatim}
## Task
Generate a memory allocation summary for the function in the system
message.
[shared allocation instructions]

## Callee Summaries
{callee_summaries}

[JSON schema]
\end{verbatim}
\end{promptbox}

\paragraph{Block prompt for large functions.}
For large functions, NLForge first summarizes code blocks before producing a full-function allocation summary. The block-level prompt is specialized to heap allocations and asks the model to suggest a descriptive pseudo-function name and signature for the block:
\begin{promptbox}{Block prompt for large functions.}
\begin{verbatim}
You are analyzing a code block from a large C/C++ function.

## Context
Function: {name}
Signature: {signature}
File: {file_path}

## Code Block
```c
{block_source}
```

## Task
Analyze this code block for heap memory allocations (malloc, calloc,
realloc, mmap, new, etc.). Also suggest a descriptive pseudo-function
name and signature for this block.

Respond in JSON with:
- suggested_name
- suggested_signature
- summary
- allocations: [type, source, size_expr, size_params, returned,
  stored_to, may_be_null]
\end{verbatim}
\end{promptbox}

This block prompt excludes ordinary stack locals and only allows \texttt{escaped\_stack} when a stack buffer escapes to a caller-visible location, which is treated as a bug.

\subsection{External Summarizer}

The external summarizer is used for libc and common POSIX or system-library functions when NLForge does not have source code for the callee. Instead of inferring behavior from function bodies, this prompt asks the model to produce memory-safety summaries from the function's well-known documented semantics.

\paragraph{Task formulation.}
The prompt identifies the target only by function name and asks the model to act as an expert in C memory-safety analysis for documented library behavior. The instantiated prompt has the following high-level form:
\begin{promptbox}{Task formulation.}
\begin{verbatim}
You are an expert in C memory safety analysis. Given the name of a
function from libc or a common POSIX/system library, provide
memory-safety summaries based on the function's well-known documented
behaviour.

Function name: {name}
\end{verbatim}
\end{promptbox}

\paragraph{JSON schema.}
The paper-level schema corresponding to the implementation prompt is:
\begin{promptbox}{JSON schema.}
\begin{verbatim}
{
  "allocation": AllocationSummary | null,
  "free": FreeSummary | null,
  "init": InitSummary | null,
  "memsafe": {
    "function": "<name>",
    "description": "<one sentence summarising overall preconditions>",
    "contracts": [
      {
        "target": "<param name>",
        "contract_kind": "disallow_null|not_freed|buffer_size|"
                         "initialized|non_negative",
        "description": "<human-readable precondition>",
        "size_expr": "<expression>",
        "relationship": "byte_count|element_count"
      }
    ]
  }
}
\end{verbatim}
\end{promptbox}

More specifically, when present, the \texttt{allocation} summary contains entries of the form:
\begin{promptbox}{JSON schema.}
\begin{verbatim}
{
  "function": "<name>",
  "description": "<one sentence>",
  "allocations": [
    {
      "type": "heap",
      "source": "<allocating primitive>",
      "size_expr": "<byte-count expression>",
      "size_params": ["<parameters affecting size>"],
      "returned": true|false,
      "stored_to": null | "param_name",
      "may_be_null": true|false
    }
  ],
  "parameters": {
    "<param>": {"role": "<role description>", "used_in_allocation": true|false}
  }
}
\end{verbatim}
\end{promptbox}

\subsection{Free Summarizer}

The free summarizer captures deallocation and resource-release behavior. Its purpose is to distinguish heap-memory deallocation from non-heap cleanup operations and to summarize how those effects are exposed at the caller-visible level.

\paragraph{JSON response schema.}
All free-summary prompt variants require JSON in the following format.
\begin{promptbox}{JSON response schema.}
\begin{verbatim}
{
  "function": "<function_name>",
  "description": "One-sentence description of what this function frees/releases",
  "frees": [
    {
      "target": "expression being freed",
      "target_kind": "parameter|field|local|return_value",
      "deallocator": "free function name",
      "conditional": true|false,
      "condition": "guard expression (omit if unconditional)",
      "nulled_after": true|false,
      "description": "for loop/transitive frees only"
    }
  ],
  "resource_releases": [
    {
      "target": "resource being released",
      "target_kind": "parameter|field|local|return_value",
      "deallocator": "close|sem_destroy|etc",
      "conditional": true|false,
      "condition": "guard expression (omit if unconditional)",
      "nulled_after": true|false
    }
  ]
}
\end{verbatim}
\end{promptbox}
If the function performs neither memory deallocation nor resource cleanup, the model is instructed to return the same schema with empty \texttt{frees} and \texttt{resource\_releases} lists and a description such as ``Does not free memory.''

\paragraph{Single-message prompt.}
The no-caching variant places the function source, metadata, callee free summaries, the shared free instructions, and the JSON schema into one prompt:
\begin{promptbox}{Single-message prompt.}
\begin{verbatim}
You are analyzing C/C++ code to generate deallocation (free) summaries.

## Function to Analyze
```c
{source}
```

Function: {name}
Signature: {signature}
File: {file_path}

## Callee Free Summaries
{callee_summaries}

## Task
Generate a deallocation summary for this function. Identify every
buffer or resource that this function frees (directly or via callees).
[shared free instructions]
[JSON schema]
\end{verbatim}
\end{promptbox}

\paragraph{Approach A: cached instructions.}
In the \texttt{cache\_mode="instructions"} configuration, the static deallocation task is cached in the system prompt while the user message contains the function source and callee summaries.
\begin{promptbox}{Approach A: cached instructions.}
\begin{verbatim}
System prompt:
You are analyzing C/C++ code to generate deallocation (free) summaries.
## Task
Generate a deallocation summary for the function provided in the user
message. Identify every buffer or resource that this function frees
(directly or via callees).
[shared free instructions]
[JSON schema with <function_name> placeholder]

User prompt:
## Function to Analyze
```c
{source}
```
Function: {name}
Signature: {signature}
File: {file_path}
## Callee Free Summaries
{callee_summaries}
\end{verbatim}
\end{promptbox}

\paragraph{Approach B: cached source.}
In the \texttt{cache\_mode="source"} configuration, the function source is cached in the system prompt and the user prompt supplies the task instructions together with the callee free summaries.
\begin{promptbox}{Approach B: cached source.}
\begin{verbatim}
## Task
Generate a deallocation summary for the function in the system message.
Identify every buffer or resource that this function frees (directly or
via callees).
[shared free instructions]

## Callee Free Summaries
{callee_summaries}

[JSON schema]
\end{verbatim}
\end{promptbox}

\paragraph{Block prompt for large functions.}
For large functions, NLForge also includes a block-level deallocation prompt that summarizes free operations before forming the full-function summary. This block prompt asks the model to infer a pseudo-function name and signature for the block in addition to the list of frees:
\begin{promptbox}{Block prompt for large functions.}
\begin{verbatim}
You are analyzing a code block from a large C/C++ function.

## Context
Function: {name}
Signature: {signature}
File: {file_path}

## Code Block
```c
{block_source}
```

## Task
Analyze this code block for free/deallocation operations. Also suggest
a descriptive pseudo-function name and signature for this block.

Respond in JSON with:
- suggested_name
- suggested_signature
- summary
- frees: [target, target_kind, deallocator, conditional,
  condition, nulled_after]
\end{verbatim}
\end{promptbox}

\subsection{Init Summarizer}

The initialization summarizer captures caller-visible post-conditions. Its role is to identify what a function definitely initializes for its caller, what range restrictions hold for caller-visible outputs, and whether the function may be \texttt{noreturn} on some or all paths.

\paragraph{JSON response schema.}
All initialization prompt variants require JSON in the following format.
\begin{promptbox}{JSON response schema.}
\begin{verbatim}
{
  "function": "<function_name>",
  "description": "One-sentence description of what this function initializes",
  "inits": [
    {
      "target": "expression being initialized",
      "target_kind": "parameter|field|return_value",
      "initializer": "how it is initialized",
      "byte_count": "concrete_expr|sizeof(T)|null",
      "conditional": true,
      "condition": "condition expression (omit if unconditional)"
    }
  ],
  "output_ranges": [
    {
      "target": "return or *out_param",
      "range": "[lower, upper] or >= 0",
      "description": "brief context"
    }
  ],
  "noreturn": false,
  "noreturn_condition": "condition under which function does not return"
}
\end{verbatim}
\end{promptbox}
If the function does not unconditionally initialize caller-visible state, the model returns the same schema with an empty \texttt{inits} list and \texttt{noreturn: false}. The implementation prompt further specifies that \texttt{output\_ranges} should be omitted when no narrower-than-type range is available, and \texttt{noreturn\_condition} should be omitted when the function always returns or is unconditionally \texttt{noreturn}.

\paragraph{Single-message prompt.}
The no-caching version combines the function source, metadata, callee initialization summaries, the shared initialization instructions, and the schema into one message:
\begin{promptbox}{Single-message prompt.}
\begin{verbatim}
You are analyzing C/C++ code to generate initialization summaries
(post-conditions).

## Function to Analyze
```c
{source}
```

Function: {name}
Signature: {signature}
File: {file_path}

## Callee Initialization Summaries
{callee_summaries}

## Task
Generate an initialization summary for this function. Identify what
this function always initializes on all exit paths -- only guaranteed,
unconditional initializations. This is a post-condition: only things
visible to the caller matter.
[shared init instructions]
[JSON schema]
\end{verbatim}
\end{promptbox}

\paragraph{Approach A: cached instructions.}
In the \texttt{cache\_mode="instructions"} configuration, the static initialization-analysis instructions are cached in the system prompt and the user message contains the function source and callee initialization summaries.
\begin{promptbox}{Approach A: cached instructions.}
\begin{verbatim}
System prompt:
You are analyzing C/C++ code to generate initialization summaries
(post-conditions).
## Task
Generate an initialization summary for the function provided in the
user message. Identify what this function always initializes on all
exit paths -- only guaranteed, unconditional initializations. This is a
post-condition: only things visible to the caller matter.
[shared init instructions]
[JSON schema with <function_name> placeholder]

User prompt:
## Function to Analyze
```c
{source}
```
Function: {name}
Signature: {signature}
File: {file_path}
## Callee Initialization Summaries
{callee_summaries}
\end{verbatim}
\end{promptbox}

\paragraph{Approach B: cached source.}
In the \texttt{cache\_mode="source"} configuration, the source is cached in the system prompt and the user prompt supplies the shared initialization instructions together with the callee summaries.
\begin{promptbox}{Approach B: cached source.}
\begin{verbatim}
## Task
Generate an initialization summary for the function in the system
message. Identify what this function always initializes on all exit
paths -- only guaranteed, unconditional initializations. This is a
post-condition: only things visible to the caller matter.
[shared init instructions]

## Callee Initialization Summaries
{callee_summaries}

[JSON schema]
\end{verbatim}
\end{promptbox}

\paragraph{Block prompt for large functions.}
For large functions, NLForge also uses a block-level initialization prompt to summarize caller-visible writes before composing a full-function summary. This prompt again asks for a pseudo-function name and signature for the block:
\begin{promptbox}{Block prompt for large functions.}
\begin{verbatim}
You are analyzing a code block from a large C/C++ function.

## Context
Function: {name}
Signature: {signature}
File: {file_path}

## Code Block
```c
{block_source}
```

## Task
Analyze this code block for initialization operations (caller-visible
only). Also suggest a descriptive pseudo-function name and signature.

Respond in JSON with:
- suggested_name
- suggested_signature
- summary
- inits: [target, target_kind, initializer, byte_count]
\end{verbatim}
\end{promptbox}

\subsection{Integer Overflow Summarizer}

The integer-overflow summarizer performs compositional value-range analysis for integer-related undefined behavior. Unlike the memory-safety passes, this prompt is explicitly restricted to arithmetic UB and is instructed not to report null dereferences, buffer overflows, use-after-free, or other memory-safety bugs handled elsewhere in the pipeline.

\paragraph{System prompt.}
At the paper level, the system prompt can be summarized as follows:
\begin{promptbox}{System prompt.}
\begin{verbatim}
You are performing value-range analysis on a C/C++ function to detect
integer-related undefined behaviour (UB). Do NOT report memory-safety
bugs.

[definitions of integer_overflow, division_by_zero, shift_ub]
[non-UB cases to ignore]
[range-tracking method]
[callee output-ranges and constraints]
[LP64 type model]
[required outputs: constraints, output_ranges, issues]

Respond with JSON only.
\end{verbatim}
\end{promptbox}

\begin{promptbox}{JSON schema.}
\begin{verbatim}
## Function
Function: {name}
Signature: {signature}
File: {file_path}

```c
{source}
```

## Callee Context
{callee_section}

## Task
Perform value-range analysis as described. Track integer ranges through
branches and arithmetic. Report UB and produce compositional
constraints / output ranges.
\end{verbatim}
\end{promptbox}

\begin{promptbox}{User prompt and JSON schema.}
\begin{verbatim}
{
  "function": "<name>",
  "description": "One-sentence summary of analysis result",
  "constraints": [
    {
      "target": "parameter name",
      "range": "[lower, upper] or descriptive",
      "description": "why this constraint is needed"
    }
  ],
  "output_ranges": [
    {
      "target": "return value or *out_param",
      "range": "[lower, upper] or descriptive",
      "description": "brief context"
    }
  ],
  "issues": [
    {
      "location": "line or expression",
      "issue_kind": "integer_overflow|division_by_zero|shift_ub",
      "description": "which operation, operand ranges, why it is UB",
      "severity": "high|medium|low"
    }
  ]
}
\end{verbatim}
\end{promptbox}

\subsection{Leak Summarizer}

The leak summarizer identifies unresolved heap allocations by combining the results of the allocation and free passes. Unlike the allocation and free prompts, this pass is explicitly comparative: it matches produced allocations against frees and then classifies any remaining obligations as either concrete leaks or caller-visible ownership that must be propagated upward.

\paragraph{Prompt structure.}
The system prompt can be summarized as:
\begin{promptbox}{Prompt structure.}
\begin{verbatim}
You are analyzing C/C++ functions for memory leaks by comparing their
allocation and free summaries.

A memory leak occurs when a heap allocation is:
- Not freed before the function returns, AND
- Not returned to the caller, AND
- Not stored to a caller-visible location.

[entry-point exception for main()]
[path-feasibility guidance]
[required outputs: leaks, simplified_allocations, simplified_frees]
\end{verbatim}
\end{promptbox}

The user prompt then provides the function source together with four pieces of analysis context: the function's own allocation summary, the function's own free summary, callee post-conditions, and an optional entry-point note.
\begin{promptbox}{Prompt structure.}
\begin{verbatim}
## Function
Function: {name}
Signature: {signature}
File: {file_path}

```c
{source}
```

## This Function's Allocation Summary
{alloc_section}

## This Function's Free Summary
{free_section}

## Callee Post-conditions
{callee_section}

## Task
Match allocations against frees.
[matching rules]
[callee propagation]
{entry_note}
\end{verbatim}
\end{promptbox}

\begin{promptbox}{JSON response schema.}
\begin{verbatim}
{
  "function": "<name>",
  "description": "One-sentence summary of leak analysis",
  "leaks": [
    {
      "allocation": "source expression (e.g., malloc(n))",
      "stored_to": "where it is stored, or null",
      "reason": "why it is not freed",
      "severity": "high|medium|low"
    }
  ],
  "simplified_allocations": [
    {
      "source": "malloc|calloc|realloc|...",
      "size_expr": "size expression or null",
      "returned": true,
      "stored_to": "field or variable, or null",
      "may_be_null": true
    }
  ],
  "simplified_frees": [
    {
      "target": "parameter or field being freed",
      "target_kind": "parameter|field",
      "deallocator": "free|custom_free|...",
      "conditional": false,
      "condition": "condition expression or null",
      "description": "optional description for transitive frees"
    }
  ]
}
\end{verbatim}
\end{promptbox}

\subsection{Memsafe Summarizer}

The memsafe summarizer infers caller obligations for memory-safe execution. In contrast to the initialization and leak passes, this prompt focuses on \emph{pre-conditions}: it asks what the caller must guarantee so that dereferences, reads, deallocations, and buffer operations in the current function are safe.

\paragraph{Task formulation.}
The single-message prompt provides the function source, metadata, and two optional context blocks: a callee-note section and an alias-context section. The callee context may be represented either as inline source annotations or as a flat list. At the implementation level, the prompt begins as follows:
\begin{promptbox}{Task formulation.}
\begin{verbatim}
You are analyzing C/C++ code to generate safety pre-condition contracts.

## Function to Analyze
```c
{source}
```

Function: {name}
Signature: {signature}
File: {file_path}

{callee_note}
{alias_context}
\end{verbatim}
\end{promptbox}

\begin{promptbox}{JSON response schema.}
\begin{verbatim}
{
  "function": "<name>",
  "description": "One-sentence summary of this function's safety requirements",
  "contracts": [
    {
      "target": "parameter or expression",
      "contract_kind": "disallow_null|allow_null|not_freed|"
                       "initialized|buffer_size",
      "description": "brief description of the requirement",
      "size_expr": "n",
      "relationship": "byte_count",
      "condition": "C expression"
    }
  ]
}
\end{verbatim}
\end{promptbox}

\begin{promptbox}{Single-message prompt.}
\begin{verbatim}
You are analyzing C/C++ code to generate safety pre-condition contracts.

## Function to Analyze
[source, signature, file]

[callee note]
[alias context]

## Task
Generate safety contracts (pre-conditions) for this function. Identify
what the caller must guarantee for this function to execute in a
memory-safe manner.

[annotation-propagation rule]
[contract-kind definitions]
[JSON schema]
\end{verbatim}
\end{promptbox}

\paragraph{Block prompt for large functions.}
For large functions, NLForge also uses a block-level memsafe prompt that summarizes pre-conditions for one code region at a time while also asking for a pseudo-function name and signature:
\begin{promptbox}{Block prompt for large functions.}
\begin{verbatim}
You are analyzing a code block from a large C/C++ function.

## Context
Function: {name}
Signature: {signature}
File: {file_path}

## Code Block
```c
{block_source}
```

## Task
Generate safety contracts (pre-conditions) for this code block. What
must the caller guarantee for memory-safe execution of this block?
Also suggest a descriptive pseudo-function name and signature.

Respond in JSON with:
- suggested_name
- suggested_signature
- summary
- contracts: [target, contract_kind, description, size_expr,
  relationship, condition]
\end{verbatim}
\end{promptbox}

\paragraph{Alternative callee-note formats.}
The implementation supports two ways of presenting callee safety information. In the annotation-based mode, the prompt includes a header indicating that callee contracts are embedded inline as \texttt{/* PRE[...] */} comments above call sites. In the flat mode, the prompt instead supplies a separate \texttt{flat\_list} under a ``Callee Safety Contracts'' header. Both encodings provide the same semantic information, but the inline form is especially useful because it localizes each callee obligation at the relevant call site.

\subsection{Verification Summarizer}
\label{sec:verifierprompt}
The verification summarizer is the final memory-safety checking pass. Unlike the earlier prompts, which infer contracts or effects, this prompt performs Hoare-style reasoning over a function body under the assumption that the function's own pre-conditions already hold. Its goal is twofold: to identify concrete memory-safety violations that remain possible under those assumptions, and to simplify the function's pre-conditions into the subset that callers must still satisfy.

\begin{promptbox}{Single-message verification prompt}
\begin{verbatim}
You are verifying memory safety of a C/C++ function using Hoare-logic-
style reasoning.

## Function Under Verification
```c
{source}
```

Function: {name}
Signature: {signature}
File: {file_path}

{type_defs_section}
## Pre-conditions (assume these hold)
{own_contracts}

{own_alloc_free_section}
## Callee Information
{callee_section}

{alias_context}
\end{verbatim}
\end{promptbox}

\paragraph{Contract simplification.}
In addition to bug finding, the prompt asks the model to simplify the function's pre-condition set. Any pre-condition that is fully discharged internally should be dropped, while requirements that callers must still satisfy should be retained in \texttt{simplified\_contracts}. This makes the verification pass both a checker and a post-processor for caller-facing contracts.

\paragraph{JSON response schema.}
The required output format is:
\begin{promptbox}{JSON response schema.}
\begin{verbatim}
{
  "function": "<name>",
  "description": "One-sentence summary",
  "simplified_contracts": [
    {
      "target": "param",
      "contract_kind": "disallow_null|allow_null|not_freed|"
                       "initialized|buffer_size",
      "description": "brief",
      "size_expr": "buffer_size only",
      "relationship": "buffer_size only"
    }
  ],
  "issues": [
    {
      "location": "line N",
      "issue_kind": "null_deref|buffer_overflow|use_after_free|"
                    "double_free|uninitialized_use|invalid_free",
      "description": "the problem",
      "severity": "high|medium|low",
      "callee": "if contract violation",
      "contract_kind": "if contract violation"
    }
  ]
}
\end{verbatim}
\end{promptbox}

\paragraph{Block prompt for large functions.}
NLForge also uses a block-level verification prompt for chunked analysis of large functions. In this version, the model receives the function context, the function's own assumed pre-conditions, and a source block rather than the full function body:
\begin{promptbox}{Block prompt for large functions.}
\begin{verbatim}
You are verifying memory safety of a code block from a large C/C++
function.

## Context
Function: {name}
Signature: {signature}
File: {file_path}

## This Function's Pre-conditions -- assume these hold
{own_contracts}

## Code Block
```c
{block_source}
```

## Task
Verify this code block for memory safety issues. Check for null
dereferences, buffer overflows, use-after-free, double-free, and
uninitialized use. Also suggest a descriptive pseudo-function name and
signature.
\end{verbatim}
\end{promptbox}

\section{Datasets}

Our evaluation uses a curated subset of the SV-COMP memory-safety benchmark suite together with the Juliet test cases included in SV-COMP. We group the selected SV-COMP tasks into four coarse families based on the dominant reasoning pattern they require: control-flow-heavy tasks, heap-manipulation tasks, array-focused tasks, and linked-list or recursive-heap tasks.

\paragraph{Control-flow subset.}
These programs depend primarily on control flow and integer reasoning rather than rich pointer-manipulation patterns.
\begin{verbatim}
# Contains programs for which the correctness depends mostly on the
# control-flow structure and integer variables. There is no particular
# focus on pointers, data structures, and concurrency.
ntdrivers-simplified/*.yml
openssl-simplified/*.yml
locks/*.yml
ntdrivers/*.yml
openssl/*.yml
memory-model/*.yml
unsignedintegeroverflow-sas23/*.yml
longjmp/*.yml
signedintegeroverflow-regression/*.yml
infeasible-control-flow/*.yml
\end{verbatim}

\paragraph{Heap subset.}
These tasks emphasize heap data structures, pointer aliasing, allocation behavior, and function-pointer reasoning.
\begin{verbatim}
# Contains tasks that require the analysis of data structures on the
# heap, pointer aliases, and function pointers.
ldv-regression/*.yml
list-ext-properties/*.yml
list-ext2-properties/*.yml
ldv-sets/*.yml
heap-data/*.yml
memsafety/*.yml
memsafety-ext/*.yml
memsafety-ext2/*.yml
memsafety-ext3/*.yml
memsafety-cve/*.yml
memsafety-bftpd/*.yml
memory-alloca/*.yml
ldv-memsafety/*.yml
ldv-memsafety-bitfields/*.yml
\end{verbatim}

\paragraph{Array subset.}
These programs require reasoning about array indices, extents, and related reachability conditions.
\begin{verbatim}
# Contains tasks for which treatment of arrays is necessary in order
# to determine reachability.
array-examples/*.yml
array-industry-pattern/*.yml
reducercommutativity/*.yml
array-tiling/*.yml
array-programs/*.yml
array-crafted/*.yml
array-multidimensional/*.yml
array-patterns/*.yml
array-cav19/*.yml
array-lopstr16/*.yml
array-fpi/*.yml
array-memsafety/*.yml
array-memsafety-realloc/*.yml
\end{verbatim}

\paragraph{Linked-list subset.}
These tasks focus on memory-safety properties of linked heap structures and list-manipulation routines.
\begin{verbatim}
# Contains tasks for checking memory safety of programs.
memsafety-broom/*.yml
heap-manipulation/*.yml
forester-heap/*.yml
list-properties/*.yml
ddv-machzwd/*.yml
list-simple/*.yml
list-ext3-properties/*.yml
\end{verbatim}

\paragraph{Juliet tests.}
In addition to the SV-COMP subsets above, we include all \texttt{juliet\_TEST} tasks distributed within SV-COMP.

\section{LLM Usage Setting}

For local inference, we host Qwen on a desktop GPU with an NVIDIA RTX 4090. The other evaluated frontier models are not run on local hardware; instead, they are accessed through their respective hosted APIs. Unless otherwise noted, inference uses provider-default settings.





\end{document}